\newif\ifpreprint 
\preprinttrue

\ifpreprint 
\documentclass[a4paper,12pt]{article}
\usepackage[cm]{fullpage}
\usepackage{amsthm}
\usepackage[affil-it]{authblk}
\else 

\documentclass[graybox,default]{svmult}

\fi 

\usepackage{type1cm}        
\usepackage{makeidx}         
\usepackage{graphicx}        
\usepackage{multicol}        
\usepackage[bottom]{footmisc}

\usepackage{newtxtext}       %
\usepackage[varvw]{newtxmath}       

\usepackage{soul}

\ifpreprint 
\usepackage{hyperref}
\hypersetup{colorlinks=true,bookmarksnumbered=true,citecolor=red,urlcolor=red}
\fi 

\makeindex             

\begin{document}
\ifpreprint 
    \title{Improving Met Office Weather and Climate Forecasts with Bespoke Multigrid Solvers}
    \author[1]{Andrew Malcolm}
    \author[2,*]{Eike~Hermann~M\"{u}ller}
    \author[3]{Robert~Scheichl}
    \affil[1]{Met Office, FitzRoy Road, Exeter EX1 3PB, UK}
    \affil[2]{Department of Mathematical Sciences, University of Bath, Bath BA2 7EX, UK}
    \affil[3]{Institute for Applied Mathematics, University of Heidelberg, Im Neuenheimer Feld 205, 69120 Heidelberg, Germany}
    \affil[*]{Email: \texttt{e.mueller@bath.ac.uk}}
    \affil[ ]{\textcopyright\; Crown Copyright, Met Office}
\else 
    \title*{Improving Met Office Weather and Climate Forecasts with Bespoke Multigrid Solvers}
    \author{Andrew Malcolm, Eike Hermann M\"{u}ller and Robert Scheichl}
    \titlerunning{Improving Met Office Forecasts with Multigrid}
    \institute{Andrew Malcolm \at Met Office, FitzRoy Road, Exeter EX1 3PB, UK \email{andy.malcolm@metoffice.co.uk} \and Eike Hermann M\"{u}ller \at Department of Mathematical Sciences, University of Bath, Bath BA2 7EX, United Kingdom \email{e.mueller@bath.ac.uk}
    \and Robert Scheichl \at Institute for Applied Mathematics, University of Heidelberg, Im Neuenheimer Feld 205, 69120 Heidelberg, Germany \email{r.scheichl@uni-heidelberg.de}\\[3ex]
    \textcopyright\; Crown Copyright, Met Office}
    %
    %
\fi
\maketitle
\ifpreprint 
    \begin{abstract}
        \noindent
        \else 
        \abstract{%
        \fi 
        At the heart of the Met Office climate and weather forecasting capabilities lies a sophisticated numerical model which solves the equations of large-scale atmospheric flow. Since this model uses semi-implicit time-stepping, it requires the repeated solution of a large sparse system of linear equations with hundreds of millions of unknowns. This is one of the computational bottlenecks of operational forecasts and efficient numerical algorithms are crucial to ensure optimal performance. We developed and implemented a bespoke multigrid solver to address this challenge. Our solver reduces the time for solving the linear system by a factor two, compared to the previously used BiCGStab method. This leads to significant improvements of overall model performance: global forecasts can be produced 10\%-15\% faster. Multigrid also avoids stagnating convergence of the iterative scheme in single precision. By allowing better utilisation of computational resources, our work has led to estimated annual cost savings of \pounds 300k for the Met Office.
        \ifpreprint 
    \end{abstract}
    \let\thefootnote\relax\footnotetext{This is a preprint of the following chapter: A.~Malcolm, E.~H.~M\"{u}ller, R.~Scheichl: \textit{Improving Met Office Weather and Climate Forecasts with
        Bespoke Multigrid Solvers}, to be published in \textit{More UK Success Stories in Industrial Mathematics}, edited by Philip J. Aston, 2024, Springer reproduced with permission of Springer Nature Switzerland AG.}
\else 
    }
\fi 

\section{Introduction}\label{sec:introduction}
Climate and weather forecasts have to be produced under very tight time constraints. For example, a typical model run to forecast the global weather for the next five days must complete within one hour, while predicting the evolution of several hundreds of millions of atmospheric variables. Efficient algorithms are crucial to achieve this goal and to make optimal use of computational resources on large, massively parallel machines. The Met Office uses the semi-implicit dynamical core codenamed ``ENDGame'' \cite{Wood2014} to predict large-scale flow. The time-stepping scheme requires the solution of a large, sparse system of linear equations in every model timestep, which can take up 25\%-60\% of the total runtime. Due to the size of the problem, direct solver algorithms are unfeasible and suitably preconditioned iterative methods have to be used. For many years the Met Office solver was based on a BiCGStab algorithm, preconditioned with a vertical line relaxation scheme, which takes into account the high aspect ratio of the grid. However, this solver suffers from large iteration counts. This becomes particularly troublesome if the solver is run in single precision: the Krylov subspace vectors can lose orthogonality and the solver might struggle to converge for tight error tolerances. To overcome this issue, we replaced  BiCGStab with a bespoke multigrid solver. Three challenges had to be overcome:
\begin{enumerate}
    \item After discretisation, the high aspect ratio of the grid results in an extremely anisotropic equation; this reduces the efficiency of standard multigrid algorithms with pointwise smoothers and uniform coarsening in all dimensions.
    \item On a latitude-longitude grid, the gridlines converge at the poles, introducing an additional, spatially varying anisotropy in the horizontal direction.
    \item Traditionally, parallel scalability of multigrid can be limited by the poor work-to-communication ratio on the coarse grids, reducing model performance.
\end{enumerate}
To overcome the first challenge, we used the tensor-product multigrid approach proposed in \cite{Boerm2001}: the grid is only coarsened in the horizontal direction, and a block-Jacobi smoother takes care of the vertical anisotropy. The second challenge was addressed with conditional semi-coarsening \cite{Buckeridge2010}: close to the poles the grid is coarsened predominantly in the latitudinal direction, thus reducing the horizontal anisotropy on the coarser grids. To address the final issue, we exploit the fact that the linear equation correlates unknowns over approximately 10 grid points, independent of the model resolution and it is sufficient to use a ``shallow'' multigrid hierarchy with 3-4 levels.

The resulting multigrid method is robust and converges rapidly. Overall, a linear solve is about twice as fast as with the previously used BiCGStab solver.
\section{Bespoke multigrid solvers}
\begin{figure}
    \begin{center}
        \includegraphics[width=0.75\linewidth]{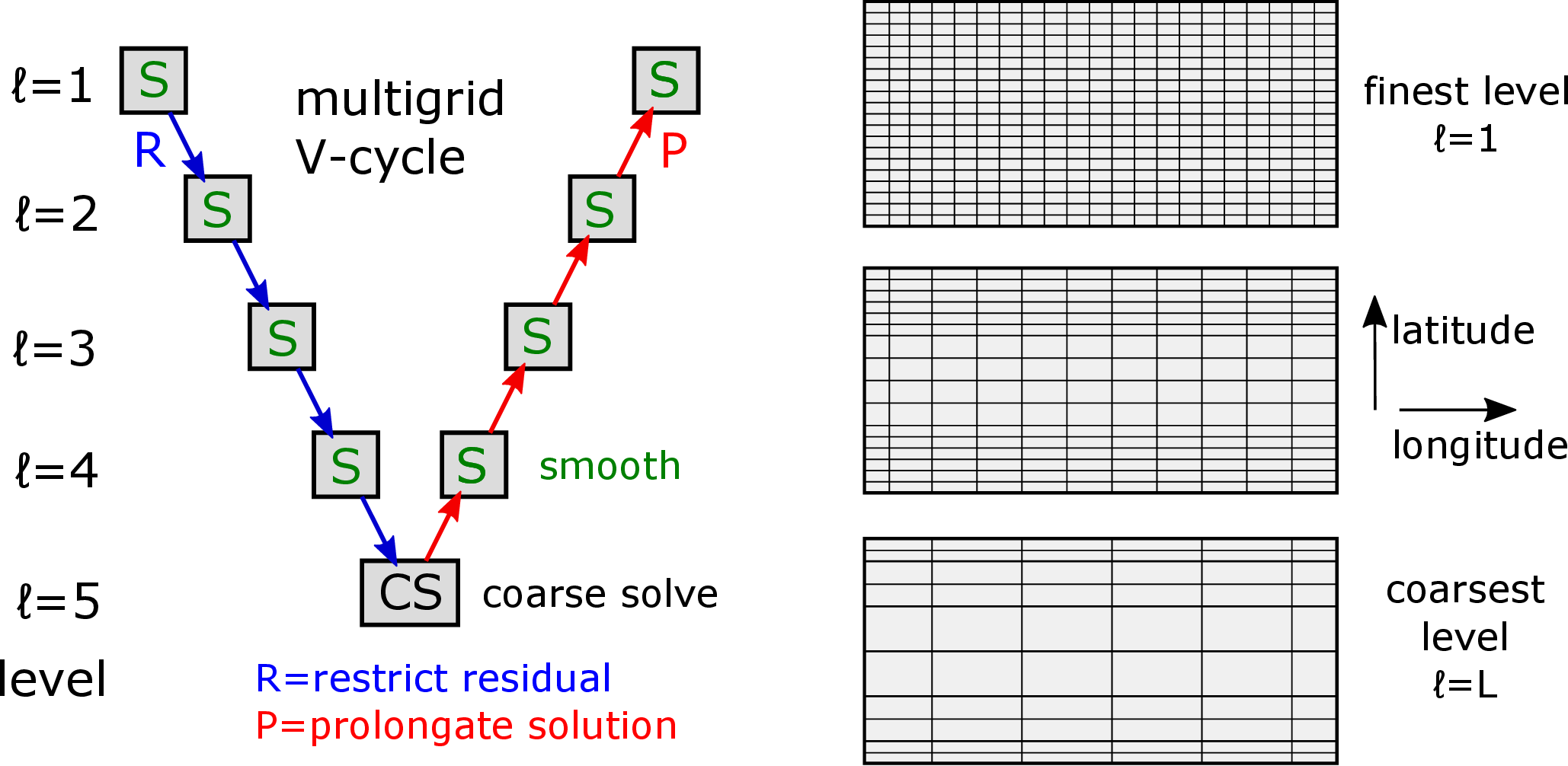}
        \caption{Multigrid V-cycle (left) and conditional horizontal semi-coarsening (right).}
    \end{center}
    \label{fig:multigrid}
\end{figure}

During each semi-implicit timestep the compressible Navier Stokes equations are solved with a quasi-Newton iteration. This in turn requires the solution of a linear system for the prognostic fields of (Exner-) pressure, wind speed, density and (potential) temperature.
By eliminating all other atmospheric variables in a Schur-complement approach, a sparse linear system, akin to a discretisation of an elliptic PDE for the pressure correction, can be obtained \cite{Wood2014}. This system has the form $Hu=b$, where $H$ is a sparse, discrete Laplacian-like matrix, $b$ is a given right-hand side and the vector $u$ represents the pressure unknowns at gridpoints. Multigrid solvers are algorithmically optimal for this type of problem, in the sense that the each iteration reduces the residual norm $||b-Hu||$ by approximately one order of magnitude. Multigrid constructs a series of coarser grids to speed up the solution of the equation. A small number of cheap smoother iterations is applied on the fine grid, which is followed by the solution of the restricted residual equation on a coarser grid, after which the coarse correction is prolongated back to the original fine grid where it is smoothed again. Applying the method recursively (Fig. \ref{fig:multigrid}), a much smaller equation is solved on the coarsest grid. The success of multigrid can be attributed to the reduction of the error at all length scales. However, the components of the algorithm, namely the smoother and coarse grid construction, have to be tuned to the problem at hand.
\subsection{Tensor product multigrid}
The elliptic operator $H=H_z+\delta H$ in ENDGame can be split into two components, where $H_z$ contains the diagonal terms and couplings within a vertical column of the grid. Since these are much larger than the horizontal couplings collected in $\delta H$, the following stationary block-Jacobi method can be used to obtain a sequence of approximations $u^{(0)},u^{(1)},\dots$ which converge to the true solution $u=H^{-1}b$:
\begin{equation}
    u^{(k+1)} = u^{(k)} - \omega H_z^{-1}\left(b-Hu^{(k)}\right)\quad\text{for some overrelaxation factor $\omega\in\mathbb{R}$}.\label{eqn:jacobi}
\end{equation}
While in the original ENDGame model the iteration in Eq. \eqref{eqn:jacobi} is used to precondition BiCGStab, here we use it as a multigrid smoother. In \cite{Boerm2001} it was shown that if this approach is combined with horizontal-only coarsening of the grid, the resulting algorithm is equivalent to a 2d multigrid method. The convergence rate is determined by the \textit{horizontal} Courant number, not by the much larger \textit{vertical} Courant number.
\subsection{Conditional semi-coarsening}
In addition to the vertical anisotropy, on a latitude-longitude grid there is also a spatially varying horizontal anisotropy $\alpha(\theta) = \Delta x_{\text{lon}}/\Delta x_{\text{lat}}$ which depends on latitude $\theta$: close to the poles the latitudinal extent $\Delta x_{\text{lat}}$ of grid cells is much larger than their longitudinal size $\Delta x_{\text{lon}}$ and this makes the smoother less efficient. To address this issue, \cite{Buckeridge2010} propose to coarsen the grid uniformly in the longitudinal direction, but make the latitudinal coarsening conditional on the anisotropy $\alpha(\theta)$, thus coarsening more agressively in this direction close to the equator. This reduces the anisotropy near the poles, which results in efficient smoothing on the coarser multigrid levels.
\subsection{Shallow multigrid}
The convergence rate of the block-Jacobi iteration in Eq. \eqref{eqn:jacobi} depends on the condition number of $H_z^{-1}H$. To a good approximation, this is given by the square of the horizontal Courant number $\nu=c\Delta t/\Delta x$ where $c$ is the speed of sound. In the Met Office model $\nu\approx 10$, independent of the resolution. Crucially, the condition number is reduced by a factor 4 each time the grid is coarsened and $\Delta x$ is doubled. This implies that after three multigrid coarsening steps the condition number is reduced to $\nu^2/4^3\approx 1.56$. Hence, it is sufficient to limit the number of multigrid levels to 4 and use a simple iterative method to solve the problem on the coarsest grid.
\subsection{Results}

\begin{figure}
    \begin{minipage}{0.48\linewidth}
        \includegraphics[width=\linewidth]{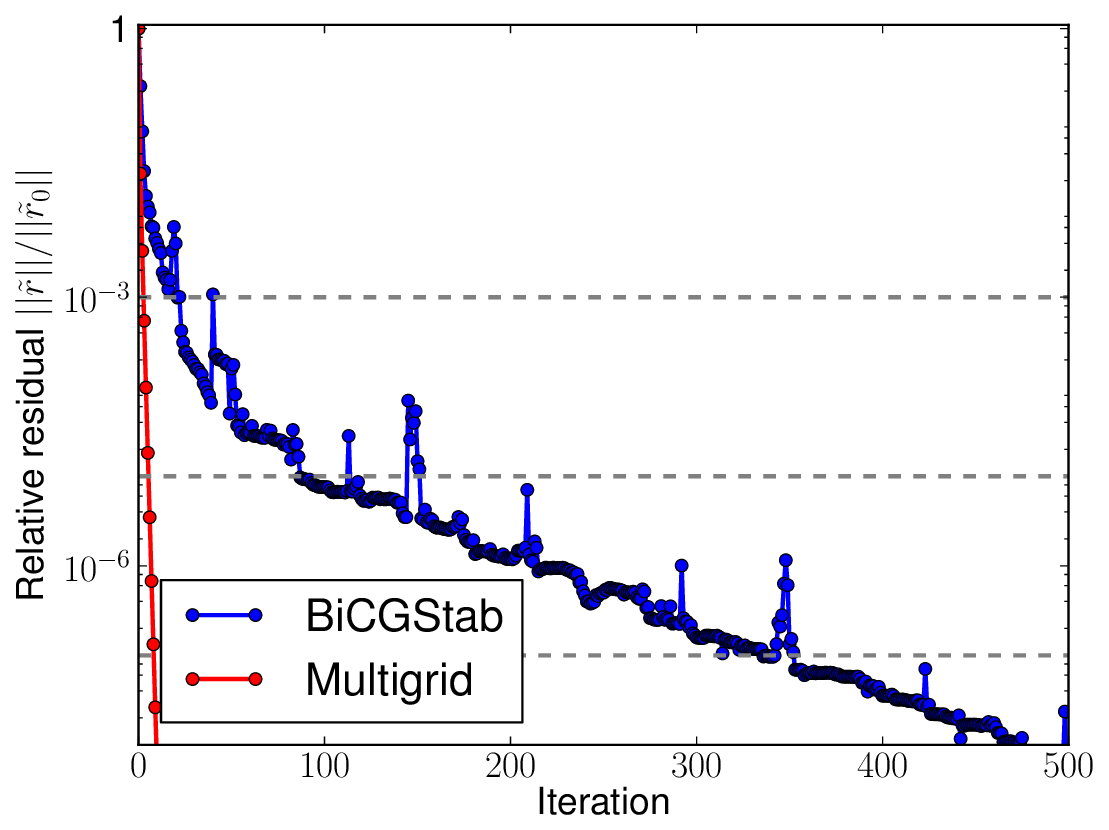}
    \end{minipage}
    \hfill
    \begin{minipage}{0.48\linewidth}
        \includegraphics[width=\linewidth]{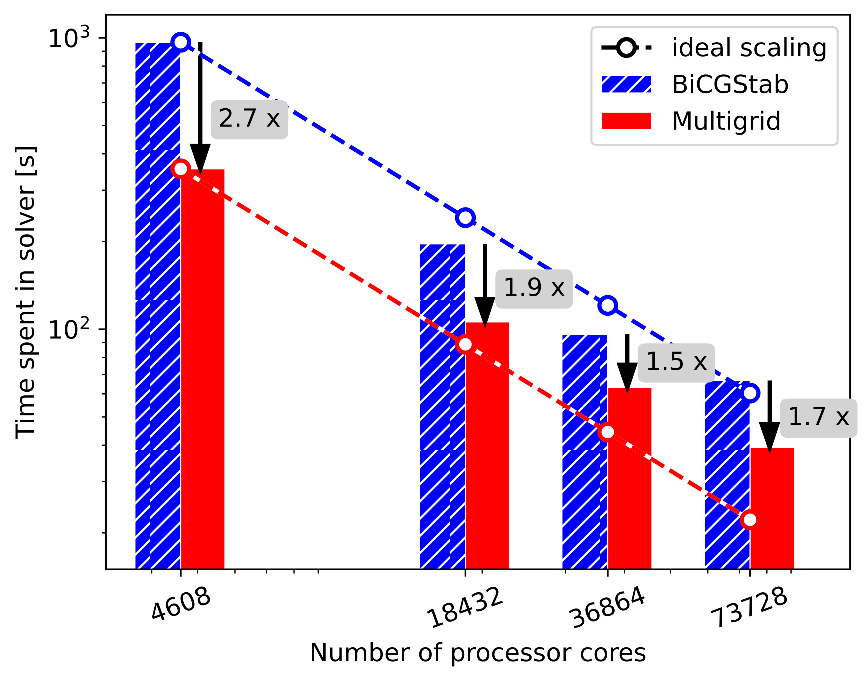}
    \end{minipage}
    \begin{center}
        \caption{Typical convergence history recorded early in the project (left) and parallel scaling of the time spent in solver for a 6km operational global model run (right).}
        \label{fig:results}
    \end{center}
\end{figure}

As Fig. \ref{fig:results} demonstrates, multigrid dramatically reduces the number of iterations and this leads to a significant acceleration of the solver. It also increases the robustness of the model, as demonstrated by the more monotonous convergence.
Tab. \ref{tab:runtimes_10km} shows the total runtime for a 75 hour deterministic forecast in the current operational configuration of the model, which corresponds to a global computational grid with $2560 \times 1920 \times 70 = 3.4\cdot10^8$ points and a spatial resolution of around $10\text{km}$ in mid-latitudes. In this case the total runtime is reduced by 8\%-16\%, while accelerating the solver itself by 29\%-39\%.
\begin{table}
    \begin{center}
        \begin{tabular}{ccccccc}
            \hline
            \# cores & \multicolumn{2}{c}{total runtime [min]} & saving    & \multicolumn{2}{c}{solver time [min]} & saving                      \\
                     & BiCGStab                                & multigrid &                                       & BiCGStab & multigrid &      \\\hline\hline
            4608     & 60.9                                    & 51.1      & 16\%                                  & 26.4     & 17.6      & 35\% \\\hline
            9216     & 32.3                                    & 28.2      & 13\%                                  & 14.3     & 10.2      & 29\% \\\hline
            11520    & 26.0                                    & 24.0      & 8\%                                   & 11.0     & 6.7       & 39\% \\\hline
        \end{tabular}
        \caption{Measured times for an operational global 10km NWP run on a $2560\times 1920\times 70$ grid.}
        \label{tab:runtimes_10km}
    \end{center}
\end{table}
\section{Achieved impact}
The key impact of this project consists in improvements to the Met Office numerical forecast products for climate and weather prediction. The code developed during the project has been used operationally since 9th December 2020. As explained in a recent news release (\url{www.metoffice.gov.uk/research/news/2020/multigrid-solver}):
\begin{quotation}The new multigrid solver will allow higher resolution forecasts to be run in the future and is already allowing better utilization of supercomputer resources.
\end{quotation}
By convincing the Met Office to adopt novel multigrid technology, the research has impacted on important decisions at the Met Office and led to significant changes to computer codes which are crucial for the organisation. According to Dr Paul Selwood, a Principal Fellow in Supercomputing and Manager of the Met Office HPC Optimisation team:
\begin{quotation}The overall investment of around 0.8 person-years (approximately \pounds 100k) of senior staff into this project demonstrates the importance the Met Office is placing on [multigrid].
\end{quotation}
To demonstrate the significance of the achieved savings in model runtime, consider the resulting monetary savings through better supercomputer utilisation. The current yearly cost of all global operational runs is \pounds 3.7mio.  As stated by Dr. Ben Shipway, Head of Dynamics Research at the Met Office:
\begin{quotation}Operational tests [of multigrid] demonstrate an average reduction of 13\% in total runtime (corresponding to significant cost savings of \pounds 300k per annum) [...] performance improvements of this order of magnitude are rare and typically only achieved once every 5 years.
\end{quotation}
The wider reach of the impact is demonstrated by the fact that the new multigrid code is currently used by one of the Met Office's international partners, the NIWA research institute in New Zealand.
\subsection{Collaboration with the Met Office}
Our research (see e.g. \cite{Mueller2014}) was carried out as part of the ``GungHo'' project. To convince the Met Office to pursue the new multigrid technology, the authors established a close working relationship with senior atmospheric scientists (Prof Nigel Wood, Dr Ben Shipway, Dr Thomas Melvin) and computational specialists (Dr Christopher Maynard, Dr Paul Selwood) through frequent mutual visits and joint meetings. To focus on the project, Eike M\"{u}ller (EM) secured a secondment through the Bath Institute of Mathematical Innovation in 2017/18. The multigrid solver was documented and integrated into the ENDGame model by the authors (2012-2019), followed by rigorous internal code review and evaluation in operational configurations (2019-2020).
\subsection{Future work}
Following the success of this project, the Met Office is committed to using multigrid technology also for its next generation model \cite{Maynard2020}. Here, multigrid has shown comparable speedups relative to Krylov subspace methods. EM has received further funding from EPSRC and the Met Office to explore the potential of multigrid for so-called hybridised discretisations, which are particular promising at higher order.
\ifpreprint 
    \section*{Acknowledgements}
\else 
    \begin{acknowledgement}
        \fi 
        This work was funded through NERC grants NE/K006754/1 and NE/J005576/1. The authors thank their collaborators at the Met Office and the members of the GungHo project. In particular, we would like to acknowledge the important contributions to the early stages of this project made by Dr Markus Gross, who tragically passed away in January 2022.
        \ifpreprint 
        \else 
    \end{acknowledgement}
\fi 
%
%
\ifpreprint 
    
\else 
    \bibliographystyle{spmpsci}
\fi 
\bibliography{paper}
\end{document}